\documentclass[preprint,12pt]{elsarticle}

\usepackage{amssymb}
\usepackage{amsmath}

\usepackage{graphicx,multirow,outlines,ulem}
\usepackage{color}
\usepackage{hhline}
\usepackage{physics}
\usepackage{hyperref}

\usepackage{caption}
\usepackage{subcaption}
\usepackage{float}
\usepackage{lipsum} 

\usepackage[mathscr]{eucal}
\journal{Physics Letters A}

\begin{document}

\begin{frontmatter}



\title{Periodicity of dynamical signatures of chaos in quantum kicked top} 


\author{Sreeram PG and M.S Santhanam } 

\affiliation{organization={Indian Institute of Science Education and Research},
            city={Pune},
            postcode={411008}, 
            state={Maharashtra},
            country={India}}
        
\begin{abstract}
A host of dynamical measures of quantum correlations -- out-of-time ordered correlators (OTOC), Loschmidt echo, generalized entanglement and observational entropy -- are useful to infer the underlying classical chaotic dynamics in quantum regime. In this work, these measures are employed to analyse quantum kicked top with kick strength $k$. It is shown that, despite the differences in their definitions, these measures are periodic with $k$, and the periodicity depends on the number of spins represented by the kicked top. The periodic behaviour arises from the structure of the kicked top Floquet operator and spans the regime in which the corresponding classical dynamics is predominantly chaotic. We also point to the reflection symmetry in Loschmidt echo and a special case of time periodicity in OTOC. This result can guide experiments towards the right choice of kick strengths to avoid repetitive dynamics.
\end{abstract}

\begin{graphicalabstract}
\end{graphicalabstract}

\begin{highlights}
\item  There is periodicity in the dynamical signatures of chaos as a function of the chaos parameter for a quantum kicked top.  Special cases of time periodicity and reflection symmetry are also presented. 
\item {Quantum kicked top is a popular experimental platform to study various quantum chaos phenomena and quantum information tasks in the laboratory. These experiments require precise control over the system and are resource-intensive since multiple measurements are required for quantification, such as extracting expectation values of an observable.  The results in this paper can guide such experiments towards choosing kick strengths to avoid repetitive dynamics. Beyond the periodicity in kick-strength, the quantum system repeats its behavior, and that regime can be carefully avoided to prevent the wastage of resources.}
\end{highlights}


\end{frontmatter}



\section{Introduction}
\label{sec1}

The manifestations of classical chaos in quantum systems is central to our evolving understanding of quantum chaotic systems. The statistical properties of spectra of quantum systems whose classical counterpart is chaotic are consistent with that of an appropriate ensemble of random matrix theory (RMT) \cite{bohigas1984characterization,berry1977level}. A good agreement between the statistics of the empirical and RMT spectrum is usually taken to imply the presence of quantum chaos. In contrast, the spectra of quantum systems whose classical limit displays regular dynamics are uncorrelated and display Poisson statistics \cite{berry1977regular}. In the last four decades, spectral statistics have been an essential tool to infer the nature of underlying classical dynamics -- whether it is regular, chaotic or of mixed type. Recently, it was shown for a quantized triangle map that RMT-type spectral correlations can arise even if the underlying classical system is only ergodic and not chaotic \cite{wang2022statistical}.

Despite the success of spectral correlations, such as the level spacing distributions as a diagnostic tool for quantum chaos \cite{chirikov1991houches,stockmann2000quantum}, they can capture only limited information about dynamics. For instance, mean level spacing can provide coarse information about typical timescales of the dynamics, but using spectral gaps to infer the time evolution of arbitrary initial states is usually not easy \cite{heller1982criteria,heller1991wavepacket}. Temporal evolution is important for the following reason. In classical systems, the nature of dynamics, whether it is chaotic or regular, can be inferred from the Lyapunov exponent, which is obtained from the time evolution of trajectories. To be on the same footing, it is desirable to characterise quantum systems using measures based on the time evolution of arbitrary initial states. 

The Out-of-Time Ordered correlator (OTOC), Loschmidt echo (LE), generalized entanglement and observational entropy (OE) are among the measures that are currently being extensively studied in the context of few- and many-body quantum chaos \cite{larkin1969quasiclassical,shenker2014black,swingle2018unscrambling,yin2021quantum,peres1984stability, jalabert2001environment,gorin2006dynamics,sreeram2023witnessing,buscemi2023observational}. Dynamical signatures of quantum chaos, such as out-of-time-ordered correlators, have emerged as an important proxy to infer the presence of exponential divergence of trajectories in the classical limit. In particular, OTOCs can distinguish between ergodicity and true chaos in classical systems \cite{wang2022statistical}. In contrast to other dynamical measures such as Loschmidt echo, an attractive feature of OTOC is that its growth rate within the Ehrenfest timescale is proportional to that of the classical Lyapunov exponent. Hence, the OTOC growth rate can be conveniently used as a ``quantum analogue'' of the classical Lyapunov exponent. 

 Loschmidt echo \cite{peres1984stability, jalabert2001environment} is closely related to the OTOC, which deals with the fidelity decay of a state subject to imperfect time reversal. Since exact time reversal is impossible in practice,  any small error amplifies rapidly and reflects in the fidelity decay when the dynamics is chaotic. 
The other two dynamical measures we consider in this paper are generalized entanglement and observational entropy. Generalized entanglement was defined in quantum systems with no natural subsystem decomposition \cite{viola2007generalized}.  Effectively, generalized entanglement is a subsystem-independent generalization which coincides with traditional entanglement when natural subsystems are considered. It was shown that generalized entanglement captures signatures of chaos in a kicked top \cite{weinstein2006generalized}. Observational entropy (OE) is the quantum analogue of the statistical entropy studied in \cite{latora1999kolmogorov}. Observational entropy was recently shown to characterise chaos in a quantum kicked top \cite{sreeram2023witnessing}. The motivation to investigate OE for signatures of chaos originates from the close connection between the coarse-grained classical entropy in Ref. \cite{latora1999kolmogorov} with the Kolmogorov-Sinai entropy rate.

 These quantities are well studied in a paradigmatic model called the quantum kicked top (QKT). It is a popular model of time-dependent quantum chaos \cite{haake1987classical, haake1991quantum,chaudhury2009quantum},
and are experimentally realised in laser-cooled atoms \cite{chaudhury2009quantum}, superconducting systems \cite{neill2016ergodic}, and NMR platform \cite{krithika2019nmr}. 

Recently, it was observed that the entanglement entropy and quantum discord, when applied to QKT, displayed periodicity with respect to the kick strength (chaos parameter) $k$, and also with respect to time in a limited sense. One consequence is that if we wish to avoid a periodic behaviour of these measures, then a constraint on the minimum number of spins comes into play. In this article, we probe if the dynamical characterisers of quantum chaos, such as OTOC, Loschmidt echo and two other entropy measures -- generalized entanglement and observational entropy -- also display periodic behaviour. The periodicities we report for these measures are different from those reported in Ref. \cite{bhosale2018periodicity} for entanglement and quantum discord. Though kicked top dynamics is obtained by the stroboscopic evolution under repeated unitary operations, periodicity is usually not seen in the evolving observables. Such periodicities are not known for quantum kicked rotor. Hence, the existence of periodicities is both a curiosity and also has experimental consequences as we describe later.


The rest of this article is organized as follows. In the next section, we provide an overview of quantum kicked top. Sections III to VI the  periodicity of dynamical signatures of chaos, starting with that of OTOC. Finally, section VII concludes by highlighting the utility of these results..

\section{Quantum kicked top and the periodicity in kick strength $k$}
Quantum kicked top (QKT) is one of the well-studied models of quantum chaos in time-dependent systems. 
 Its quantum dynamics can be conveniently studied through the period-1 Floquet operator given as
\begin{equation}
U(k) = \exp (-i \frac{k}{2j} J_z^2 ) \exp(-i \alpha  J_y). 
\label{eq:uniqkt}
\end{equation}
This unitary involves precession about the $y$-axis by an angle $\alpha$ followed by a torsion proportional to $J_z$ about $z$-axis. The parameter $k$ is the strength of the kick~(torsion), and $j$ is the angular momentum quantum number.

The action of this unitary operator on a quantum state $\ket{\psi(0)}$ describes its evolution such that the evolved wave function after $m$ time steps is
\begin{equation}
\ket{\psi(m)}= U^m(k) \ket{\psi(0)}.
\label{eq:tevolve}
\end{equation}
The evolution of the angular momentum operator can be seen in the Heisenberg picture as
\begin{equation}
J_i^\prime = U^\dagger(k) ~ J_i ~ U(k), \;\;\;\; i=x,y,z.
\end{equation}
 Chaos measures  we talked about in the introduction are expected to capture the properties of the underlying dynamics.  Therefore, if there is a periodicity in the parameter space of the unitary Floquet, it must be reflected in all of them. Let us probe if such a periodicity in kick strength $k$ arises in the case of the kicked top unitary by taking   $k \to k + \kappa_j$, where $\kappa_j$ is a positive real number. The unitary operator in Eq. \ref{eq:uniqkt} under this transformation becomes
\begin{equation}
\begin{split}
     U({k+ \kappa_j}) & = \exp (-i \frac{k+\kappa_j}{2j} J_z^2 ) ~ \exp(-i \alpha  J_y), \\ 
                         & = \exp(-i \frac{\kappa_j}{2 j} J_z^2)  ~ U(k).
\end{split}
\label{eq:optrans}
\end{equation}
$U$ is periodic in $k$  up to a uni-modular constant $z$  if the unitary pre-factor

\begin{equation}
\exp(-i \frac{\kappa_j}{2 j} j_z^2) = z ~ \mathcal{I}_d,
\label{eq:invcond}
\end{equation}
where $z$ is some complex number with $|z|=1,$ and $\mathcal{I}_d$ is the identity operator. The periodicity in $k$ is that of $\kappa_j$ if Eq. \ref{eq:invcond} holds. Note that $z~ \mathcal{I}_d$ commutes with any operator. Since quantum evolution involves unitary conjugation, the constant pre-factor $z$  multiplies as $z^*z=1$, and  results in perfect periodicity of the quantum chaos measures as we will see.

What are the values of $\kappa_j$ for which periodicity holds? Equation \ref{eq:invcond} can be satisfied for positive integer values of $j$ if $\kappa_j = 4p\pi j.$ In this case the pre-factor becomes
\begin{equation}
\exp(-i 2 p \pi j_z^2)= \mathcal{I}_d, \;\;\;\;  p=1,2,3, \dots .
\label{eq:even_period}
\end{equation}
The condition in Eq. \ref{eq:invcond} can also be satisfied for positive half-integer values of $j$ if $\kappa_j=2p\pi j$ so that
\begin{equation}
     \exp(-i  p \pi j_z^2) =\frac{1}{\sqrt{2}}(1-i) \mathcal{I}_d. 
\label{eq:odd_period}
\end{equation}
\section {OTOC for kicked top} 
\label{sec 3}
Out-of-time-ordered correlators (OTOC) have emerged as a popular measure of quantum chaos
\cite{larkin1969quasiclassical,shenker2014black,swingle2018unscrambling}. For chaotic systems, they display a quantum analogue of the butterfly effect, a classic feature of chaos.  
We work with the infinite temperature OTOC, defined as
\begin{equation}
    C(t) = -\frac{1}{2} \Tr(\frac{I}{d}[W(t),W]^2 ).
    \label{otoc1}
\end{equation}
 where  $W$ is a local operator acting in a small subspace of the system. The operator $W$ is evolved under the kicked top Floquet: $W(t)= U(t) W U^\dagger (t)$.
To study the OTOC of the kicked top unitary in Eq. \ref{eq:uniqkt}, we evolve an arbitrary operator $W=W(0)$  stroboscopically for $m$ iterations using
\begin{equation}
W(m;k) = U^m(k) ~ W ~ {U^\dagger}^m(k), \;\;\; m=1,2,3, \dots ,
\end{equation}
where $k$ is the strength of the kick applied to the top. In terms of Eq. \ref{otoc1}, our object of interest is the time-evolving OTOC given by
\begin{equation}
    C(m; k) = \frac{-1}{2} \Tr(\frac{I}{2j+1} \left[  W(m;k) , W \right]^2 ).
    \label{eq:otoc_kt}
\end{equation}


\subsection{Periodicity of OTOC with kick strength $k$}
The central result we shall discuss below is that, for the kicked top system, measures such as OTOC, Loschmidt echo, generalized entanglement and observational entropy display periodic behaviour as a function of kick strength $k$. 
Let us begin by analyzing the occurrence of periodicities in OTOC under a change of parameter, i.e., we are seeking invariance of OTOC under the transformation $k \to k + \kappa_j$ for some periodicity $\kappa_j$. 
To realize periodicity, let us evolve the operator $W$ in OTOC as follows:
\begin{align}
    W(m & ; k+\kappa_j) =  U^m(k+\kappa_j) ~ W ~ {U^{\dagger}}^m(k+\kappa_j) \nonumber \\
                    & =  \left[ \exp(-i \frac{\kappa_j}{2 j} j_z^2)  ~ U(k) \right]^m W \left[ U^{\dagger} ~ \exp(i \frac{\kappa_j}{2 j} j_z^2) \right]^m \nonumber  \\
                    & = U^m(k) ~ W ~ {U^{\dagger}}^m(k) = W(m;k)
\end{align}
The last equality is obtained by using the condition in Eq. \ref{eq:invcond}.
Thus, using Eqs. \ref{eq:even_period}-\ref{eq:odd_period}, the central result can be stated as 
\begin{equation}
\begin{split}
    C(m;k) & = C(m; k + \kappa_j), \;\; \kappa_j=4p\pi j, \;\; j \in \mathbb{Z}^+\\
           & = C(m; k + \kappa_j), \;\; \kappa_j=2p\pi j,  \;\; j \in \mathbb{Z}^+-\frac{1}{2},
    \end{split}
    \label{eq:otoc-period1}
\end{equation}
where $\mathbb{Z}^+$ denotes the set of positive integers and $\mathbb{Z}^+-\frac{1}{2}$ is the set of positive half-integers. At any fixed time step $m$, we can recognise two distinct periodicities depending on $j$. Note that this argument does not involve $J_y$ operator in Eq. \ref{eq:uniqkt} and $\kappa_j$ is valid for any value of $\alpha$.

\begin{figure}[t]
\centering
\includegraphics*[width=0.8\textwidth]{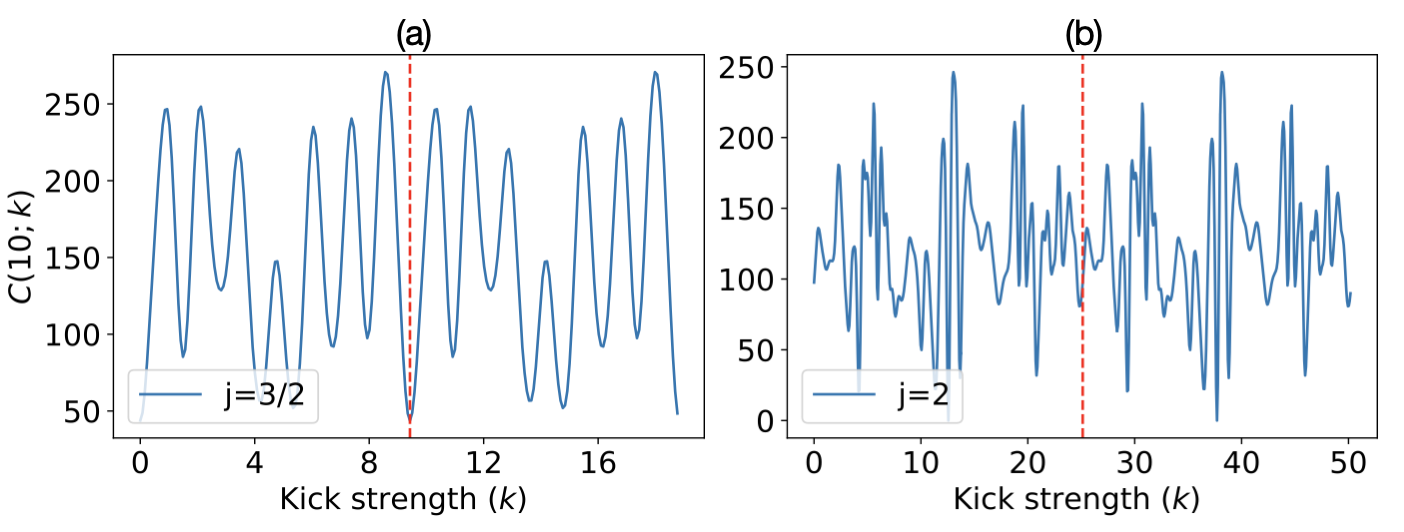}
 \caption{
 The figure shows the $k-$periodicity in $C(m=10;k)$ generated by a random Hermitian operator $W$ at $\alpha=\pi/4$ for (c) $j=3/2$, the red (dashed) line at $ \kappa_j=2\pi j=3 \pi \approx 9.42,$ marks one period, and (d) $j=2$, (dashed) line at $\kappa_j=4\pi j=8 \pi \approx 25.13,$ marks one period.}
 \label{fig:period_random_op}
\end{figure}

\begin{figure}[t]
\centering
\includegraphics*[width=0.8\textwidth]{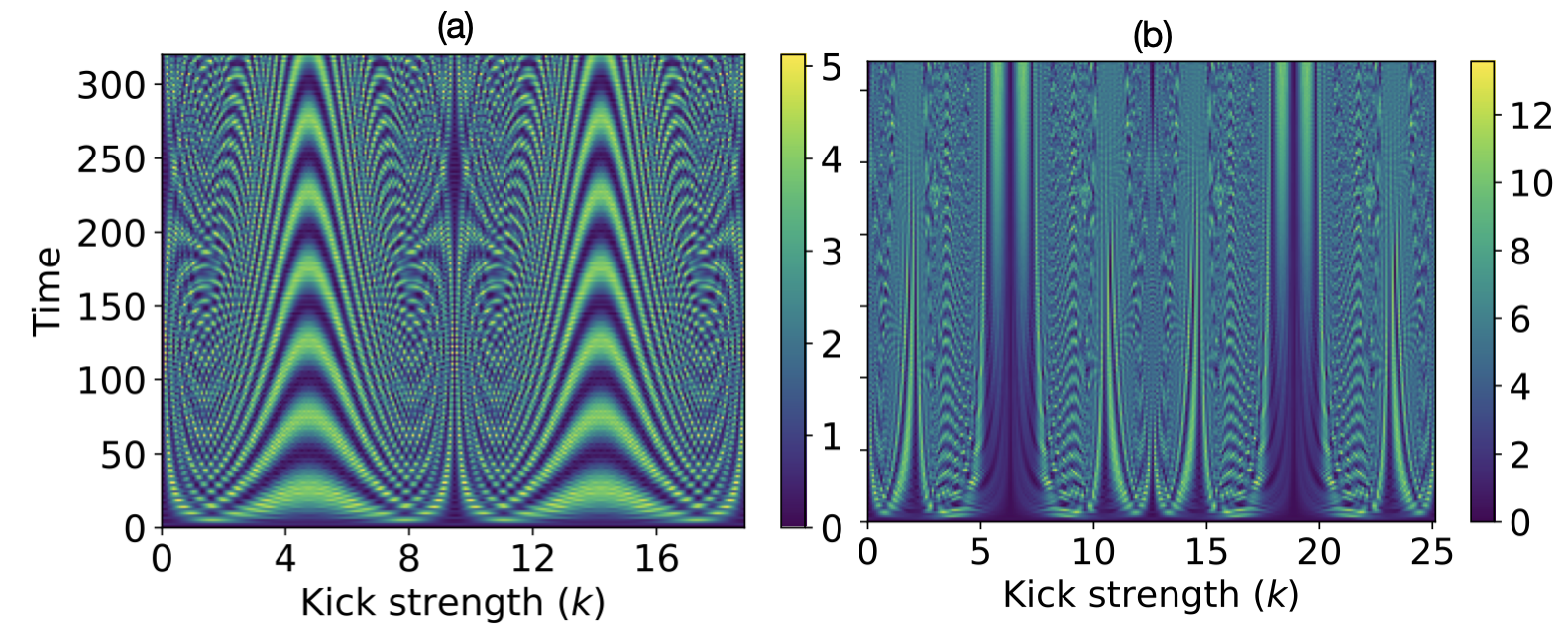}
 \caption{Image plot of $C(m;k)$ with $\alpha=\pi/4$. (a) For $j=3/2$, periodicity of $\kappa_j=2\pi j$ occurs for all $m$. (b) For $j=2$, periodicity of $\kappa_j=4\pi j$ occurs for all $m$.  }
 \label{fig:period_otoc_anytime}
\end{figure}

To demonstrate this periodicity through simulations, we have chosen $W$ to be a random matrix drawn from the Gaussian orthogonal ensemble (GOE) of random matrix theory. 
As seen in Fig. \ref{fig:period_random_op}, $C(10;k)$ as a function of $k$ displays a periodicity of $4\pi j$ as anticipated by Eq. \ref{eq:otoc-period1} with $p=2$. The seemingly random pattern of the OTOC $C(m;k)$ in Fig. \ref{fig:period_random_op} (as a function of $k$)  depends on the choice of the operator $W.$ Each different choice of $W$ gives rise to a different random-looking but periodic pattern. Irrespective of the choice of $W,$ Eq. \ref{eq:otoc-period1} guarantees that OTOC will always display periodicity with $k$ as obtained in Eq. \ref{eq:otoc-period1}. Although we choose $m=10$ in Fig. \ref{fig:period_random_op} for illustration, the periodicity in $k$ is independent of the time. To show this, we plot an OTOC for an extended time in Fig.  \ref{fig:period_otoc_anytime}. The same periodicities mentioned above can be identified from the figure.

 We also point out that  $C(m;k)$ generally does not show any time periodicity (see Sec. \ref{time_period_of_otoc} for an exception), as seen in Fig. \ref{fig:period_otoc_anytime}. Only a finite time snapshot is shown for clarity. It turns out that the combination $k=\pi/40$ and $j=1$ is an exceptional case which shows periodicity in time for rational values of $k$ as we will discuss later.

\begin{figure}
    \centering
    \includegraphics*[width=0.6\textwidth]{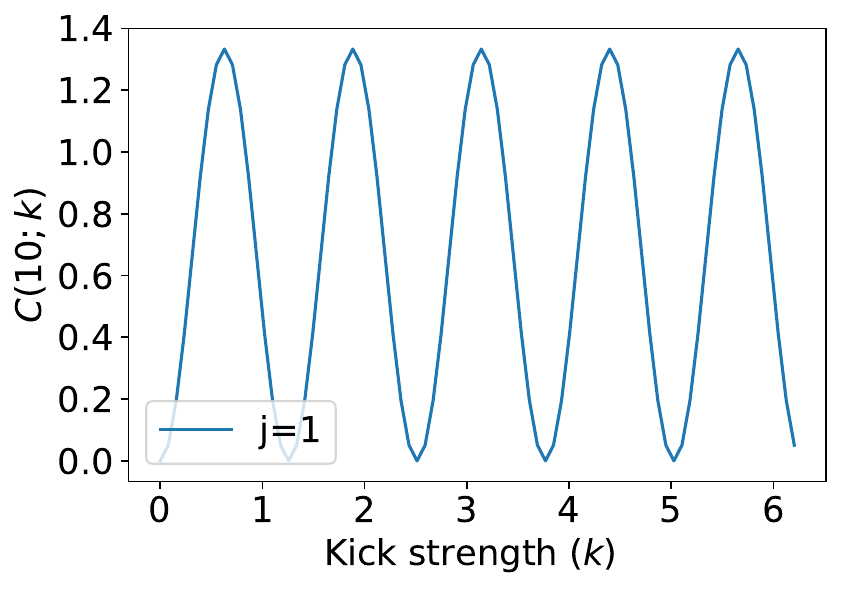}
    \caption{OTOC $C(10;k)=-\frac{1}{2} \Tr(\frac{I}{3}[J_z(10),J_z(0)]^2 )$ as a function of $k$ for $j=1, \: \alpha= \pi/2$. Note that periodicity is smaller than $2\pi j$.}
    \label{fig:jz}
\end{figure}

Finally, we note that periodicities smaller than $4\pi j$ for $j \in \mathbb{Z}^+ $ are observed in some special cases. One such special case is illustrated in Fig. \ref{fig:jz} for the OTOC given by $-\frac{1}{2} \Tr(\frac{I}{d}[J_z(m),J_z(0)]^2 )$. Further, it is seen that if $\alpha \neq \pi/2$, a periodicity of $\kappa_j=4 \pi j $ is observed for $j\in \mathbb{Z}^+$ regardless of $m$ and the operators constituting the OTOC. For  $j \in \mathbb{Z}^+-\frac{1}{2}$, periodicities smaller than $2\pi j$ is not observed regardless of the choice of operators, $m$ and $\alpha.$


\subsection{Special cases of time periodicity} \label{time_period_of_otoc}
In this subsection, we study the periodicity in $C(m;k)$ with $j=1$ and $\alpha=n\pi$ ($n \in \mathbb{Z}$). The unitary operator $U(k)$ with $n=1$ and $\alpha=\pi$ has a rather simple form: 
\begin{equation}
    U(k)=\left(\begin{array}{ccc}0&0&e^{-ik/2}\\
          0&-1&0  \\
         e^{ik/2} &0&0 
     \end{array} \right).
\end{equation}
The operator for after $m$ iterations yields
\begin{equation}
    U^m(k)=\left(\begin{array}{ccc}\alpha_m&0&\beta_m\\
          0&(-1)^m&0  \\
        \beta_m &0&\alpha_m 
     \end{array} \right),
     \label{um_pi}
\end{equation}
where $\alpha_m=\frac{1}{2}\left((e^{-\frac{ik} {2}})^m+(-e^{-\frac{ik} {2}})^m \right)$ and $\beta_m=\frac{1}{2}\left((e^{-\frac{ik} {2}})^m-(-e^{-\frac{ik} {2}})^m \right)$. The structure of these matrix elements is such that for any integral $m$, either $\alpha_m=0$ or $\beta_m=0$. Therefore, we always get $U^m(k)$ to be a diagonal or an anti-diagonal matrix with two corner matrix elements being $u_{1,2j+1}=u_{2j+1,1} \ne 0$. 

This special structure of the unitary leads to a time periodicity of OTOC. When $k=r \pi/s, \:(r,s \in \mathbb{Z})$, we find a time period of $4s$ for odd $r$,  and $2s$ for even $r$ values. This is easily seen from the functional form of $\alpha_m$ and $\beta_m$. For instance, when $m$ is even and $k=r\pi/s$, we have that  $\alpha_m= \cos(k m/2)= \cos(r\pi m/2s)$. Now, 
\begin{equation}
    \alpha_{m+4s}= \cos(\frac{r\pi (m+4s)}{2s})=\cos(\frac{k m}{2}+2r\pi).
\end{equation}
Since $r $ is an integer, $\cos(\frac{k m}{2}+2r\pi)= \cos(\frac{k m}{2}),$ and therefore $\alpha_m=\alpha_{m+4s}.$ When $r$ is an even integer, one can easily verify that $\alpha_{m+2s}=\alpha_m$. The same arguments follow when $m$ is an odd integer and for $\beta_m$. We have numerically verified  this periodicity by choosing random operators for OTOCs.

\section{Periodicity in Loschmidt Echo}
Loschmidt echo (LE) is another well-known dynamical measure of quantum chaos \cite{peres1984stability,gorin2006dynamics}.
LE is defined as the overlap between two slightly different quantum evolutions.

The state-independent definition of LE is given by \cite{zanardi2004purity,garcia2016lyapunov}
\begin{equation}
      F(m;k,k')= \frac{1}{d(d+1)} \left[ d+ \left|\Tr(U^{-m}(k')U^m(k)) \right|^2 \right],
     \label{avg_le}
\end{equation}
where forward time evolution with the kicked top unitary depends on kick strength $k$, while the time-reversed dynamics depends on $k'$. Time is denoted by $m.$ For classically chaotic systems, LE shows rapid decay as a function of time and the decay rate depends on the Lyapunov exponent in certain regimes. { For smaller perturbations, however, the decay is Gaussian}.


\begin{figure}
     \centering
\includegraphics[width=0.8\textwidth]{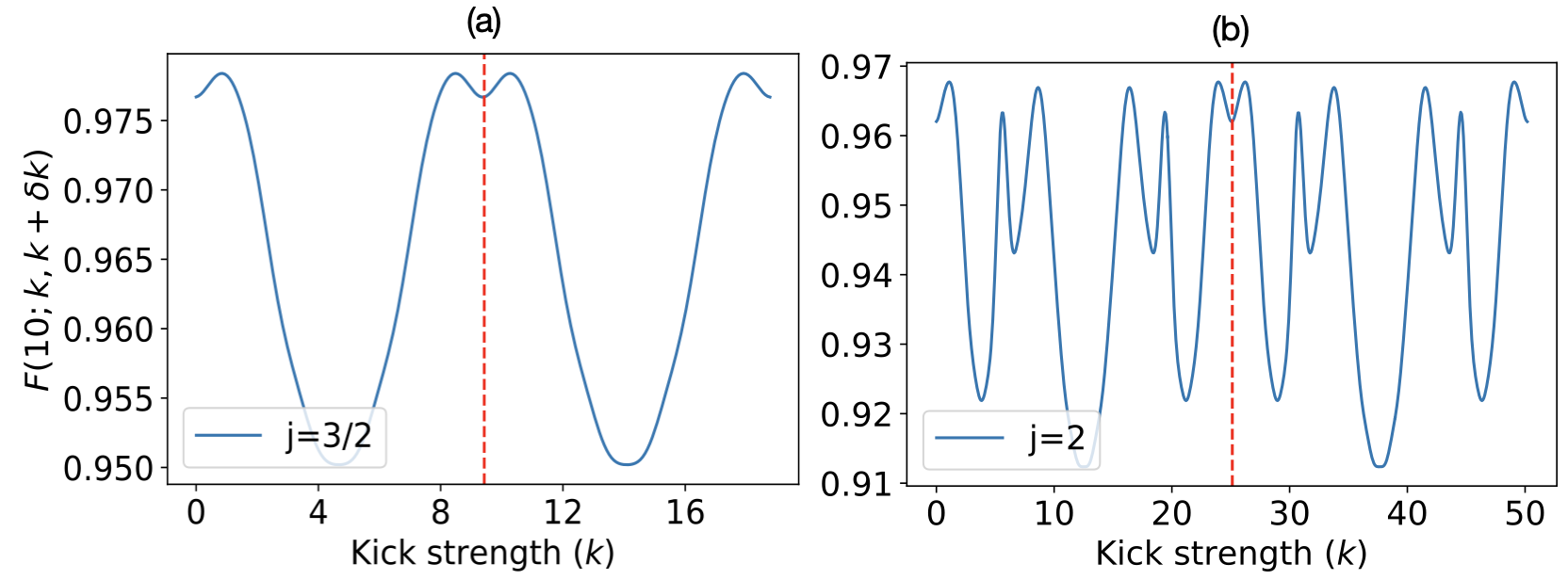}
    \caption{ (a) Periodicity of the Loschmidt echo for $j=3/2$, $\alpha=\pi/4, \: m=10$ and $k'=k+0.1.$ The red vertical line is at $ \kappa_{3/2}=2\pi j=3 \pi \approx 9.42,$ marking one period. (b) Periodicity of the Loschmidt echo for $j=2$, $\alpha=\pi/4, \: m=10$ and $k'=k+0.1.$  The red vertical line is at $\kappa_{2}=4\pi j=8 \pi \approx 25.13,$ marking one period. }
    \label{fig:LE_period_alpha_piby4}
\end{figure}

         
 %

To explore the periodicity of LE in $k$, we assume that $k'=k+\delta k$ in Eq. \ref{avg_le} and demand that  
\begin{equation}
     F(m;k,k + \delta k)=  F(m;k+\kappa_j, k+\kappa_j+\delta k)
    \label{eq:Fperiod}
\end{equation}
should hold for all fixed values of integer times $m$. It is straightforward to show that Eq. \ref{eq:Fperiod} can be satisfied by periodicity $\kappa_j$ identical to that for OTOC :
\begin{equation}
\begin{split}
    \kappa_j & = 4p \pi j, \;\;\; j \in \mathbb{Z}^+\\
           & = 2p \pi j, \;\;\; j \in \mathbb{Z}^+-\frac{1}{2}.
    \end{split}
    \label{eq:le-period1}
\end{equation}
Independent of other parameters. Figure \ref{fig:LE_period_alpha_piby4} displays the numerically simulated echo for kicked top system, confirming the periodicities predicted by Eq. \ref{eq:le-period1}. We would like to emphasise that the period $\kappa_j$ is also independent of the amount of perturbation, $\delta k.$

Like  OTOC,  Loschmidt echo also displays $\kappa_j$ smaller than allowed by Eq. \ref{eq:le-period1} but without violating the general rules in Eq. \ref{eq:le-period1}. For instance, when $\alpha=\pi/2$, Eq. \ref{avg_le} yields a period of $2\pi j$ when $j \in \mathbb{Z}^+-\frac{1}{2}$, and $\pi j$, when $j\in \mathbb{Z}^+$. These can be verified using the analytical expressions in Ref. \cite{sreeram2021out}. For $j=3/2$ and $\alpha= \pi/2$, Eq. \ref{avg_le} gives
\begin{equation}
     F(m;k,k') = |\alpha_m \tilde \alpha_m^*+ \beta_m \tilde \beta_m^*+ \beta_m^* \tilde \beta_m+\alpha_m^* \tilde \alpha_m|^2,  \label{eq:le}
\end{equation}
\begin{equation}
\begin{split}
    \alpha_m &= T_m (\chi)+ \frac{i}{2} U_{m-1}(\chi) \cos(k/3)\\
    \beta_m &= \frac{\sqrt{3}}{2} U_{m-1}(\chi) \exp(ik/3),
\end{split} \label{eq:alpha_beta}
\end{equation}
where $\chi= \sin(k/3)/2$, $T_m$ and $U_m$ are Chebychev polynomials of the first and second kind, respectively. They are defined as
\begin{equation}
    T_m(\cos \theta )= \cos(m\theta), \:\; U_m(\cos \theta)= \frac{\sin((m+1)\theta)}{\sin \theta}.
    \nonumber
\end{equation}
In Eq. \ref{eq:alpha_beta}, $\chi=\cos \theta=\sin(k/3)/2$. Substituting $k \rightarrow k+2 \pi j= k+3 \pi,$ and  $k' \rightarrow k'+3 \pi,$
we get
\begin{equation}
\alpha_m =
    \begin{cases} 
        \alpha_m^*, &  m \: \textrm{even},\\
         -\alpha_m^*, & m \: \textrm{odd}, 
    \end{cases} \label{eq:alpha_p}
\end{equation}
\begin{equation}
\beta_m =
    \begin{cases} 
        \beta_m, &  m \: \textrm{even},\\
         -\beta_m, & m \: \textrm{odd}. 
    \end{cases} \label{eq: beta_p}
\end{equation}
Substituting these back in Eq. \ref{eq:le}, it is easy to see that $  F(m;k,k')$ remains invariant for all $m$. Therefore, $F(m;k,k')= F(m;k +2 \pi j,k'+2 \pi j)$ for all $k, k'\: \in (0,2 \pi j).$  Any smaller translation of $k$ than by $2 \pi j$ will not have this property. 

Similarly, one can verify the periodicity for $j=2.$ Eq. \ref{avg_le} leads to \cite{sreeram2021out}
\begin{equation}
\begin{split}
     F(m;k,k')=\frac{1}{30} \left( 5+|1+ e^{im \delta k/4}(\alpha_m \tilde \alpha_m^*+ \right. \\ \beta_m \tilde \beta_m^*+ \beta_m^* \tilde \beta_m+\alpha_m^* \tilde \alpha_m +2 e^{3im \delta k/4}\cos^2(m \pi/2)\\+ \left. \sin^2(m\pi/2 ) \cos^2(3 \delta k/8)|^2 \right),
   \end{split}
\end{equation}
where $\delta k$ refers to the strength of the perturbation, which hampers the time reversal \cite{sreeram2021out}. For $j=2,$ the terms $\alpha_m$ and $\beta_m$ take the form
\begin{equation}
\begin{split}
    \alpha_m &= T_m (\chi)+ \frac{i}{2} U_{m-1}(\chi) \cos(k/2)\\
    \beta_m &= \frac{\sqrt{3}}{2} U_{m-1}(\chi) \exp(ik/2),
\end{split} \label{eq:alpha_beta_j2}
\end{equation}
where $\chi= \sin(k/2)/2$.
Taking $k \rightarrow k+ \pi j= k+2\pi$, and $k' \rightarrow k'+ 2\pi $, we get the same conditions as in Eq. \ref{eq:alpha_p} and \ref{eq: beta_p}. Under those transformations, the function $(\alpha_m \tilde \alpha_m^*+   \beta_m \tilde \beta_m^*+ \beta_m^* \tilde \beta_m+\alpha_m^* \tilde \alpha_m )$ remains invariant, and hence the function $F(k,k',m).$
\subsection{Reflection symmetry of Loschmidt echo}
In addition to the periodicity in kick strength, we point out the existence of a reflection symmetry about $\kappa_j/2$ for Loschmidt echo, which can be observed in Fig. \ref{fig:LE_period_alpha_piby4}. For $j \in \mathbb{Z}^+$, and a set of the chaos parameters $k_1, k_1' \in (0,2 \pi j)$,  this symmetry arises due to the existence of simple relationships of the form
\begin{equation}
     F(m;k_1, ~k_1') =  F(m;4\pi j - k_1, ~ 4 \pi j - k_1').
\label{eq:alpha_beta_j2}
\end{equation}
Similarly, for half-integer $j$, a reflection symmetry about $k=\pi j$ exists. This symmetry is independent of $\alpha$ and  $\delta k$. This reflection symmetry is similar to that reported for several quantum correlations such as entanglement and quantum discord \cite{bhosale2018periodicity}. 

\section{Periodicity in Generalized entanglement}
Generalized entanglement (GE) was introduced as a subsystem-independent generalization of entanglement \cite{barnum2004subsystem} when the system does not have a natural tensor product structure.
Generalizing the von Neumann entanglement entropy, one can define purity with respect to a set of \textit{distinguished observables} without referring to subsystems.  
The set of distinguished observables constitutes the limited means of measuring the system, analogous to a subsystem observer, who can only measure local observables. If the system has a natural subsystem decomposition, then one can get back standard entanglement by choosing the set of all uni-local observables  corresponding to a subsystem \cite{viola2007generalized}. 

Let $\ket{\psi_0}$ denote a pure state in Hilbert space $\mathcal H$. Using a chosen set of observables $\Omega = \lbrace \mathcal{A}_l\rbrace_{l=1}^L$, generalized purity by measurement can be defined as
\begin{equation}
    P_\Omega({\psi_0})= K \sum_l |\bra{\psi_0} \mathcal{A}_l \ket{\psi_0}|^2,
\end{equation}
where $K$ normalizes maximum purity to one. Then, with respect to the set $\Omega$ generalized entanglement is defined as $\mathrm{GE}_\Omega= 1-P_\Omega$.  Physically, GE quantifies the extent of a state with respect to a set of operators \cite{weinstein2006generalized}. The extent of states in different directions can distinguish a regular motion from a chaotic one \cite{cerdeira1991quantum,peres1997quantum}. 
GE is established as a signature of chaos in quantum kicked top \cite{weinstein2006generalized}. 
\begin{figure}
     \centering
    \includegraphics[width=0.8\textwidth]{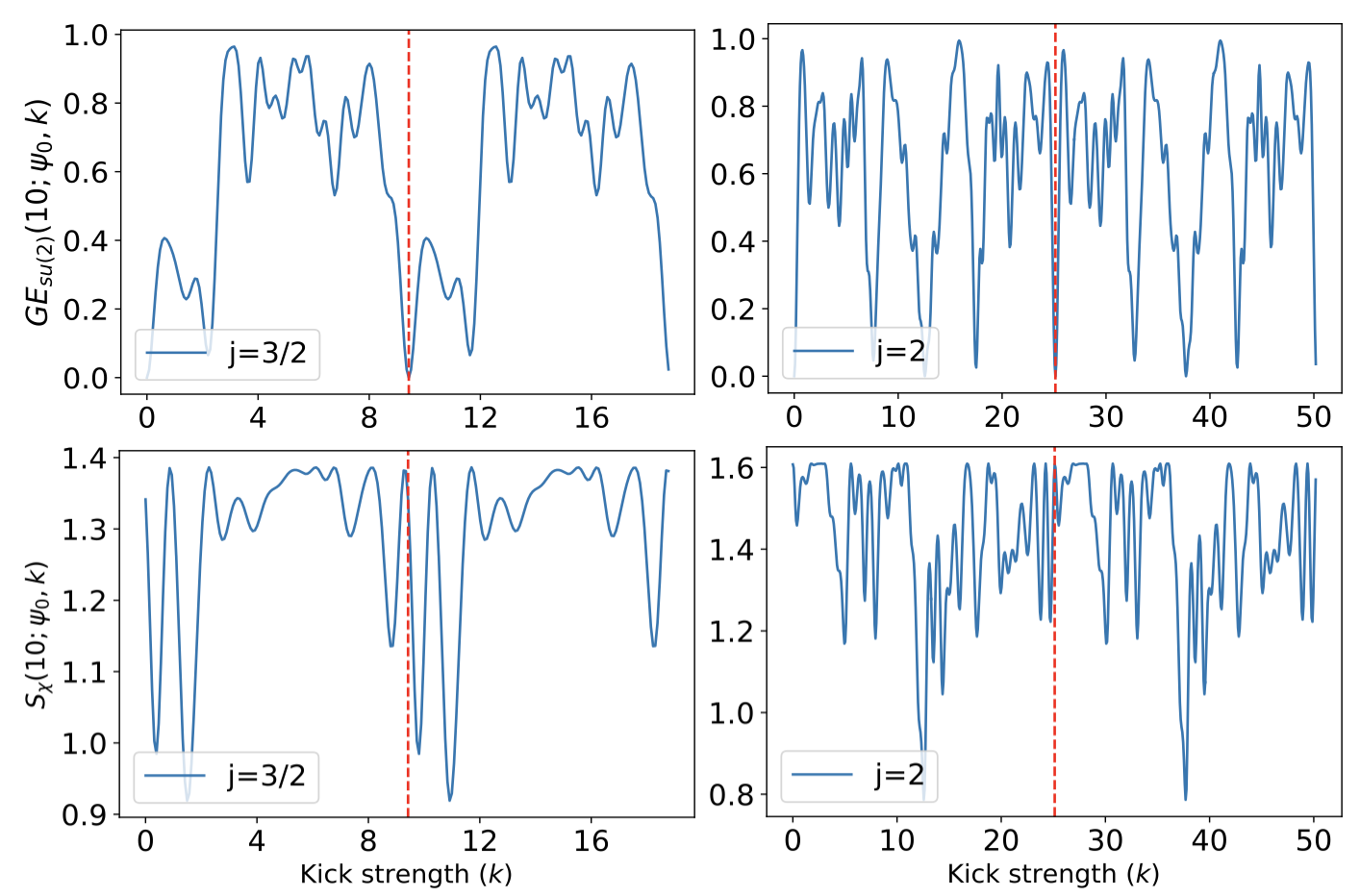}
    \caption{Top panel shows the periodicity of GE, and the bottom panel that of OE with $\alpha=\pi/4$ for (a) $j=3/2$ and $m=10$,  the red (dashed) line marks the expected periodicity based on Eq. \ref{eq:ge-period1} drawn at $\kappa_{3/2}=2\pi j=3 \pi \approx 9.42$. (b) for $j=2$ and $m=10$, the red (dashed) line is drawn at $\kappa_2=4\pi j=8 \pi \approx 25.13$. }
   \label{fig:period_GE}
\end{figure}

To study GE, we choose the initial state to be a  spin-coherent state at time $m=0$
\begin{equation}
 \ket{\psi(m=0;\theta,\phi)} = \psi_0 = \frac{e^{\beta J_-}}{(1+\beta \beta^*)^j } \ket{j,j},
 \label{eq:coh} 
\end{equation}
parameterized by $(\theta,\phi)$, and chosen uniformly at random. In this, $\beta = e^{i\phi} \tan (\theta/2), J_- = J_x -i J_y$, and $\ket{j,m}$ is the joint eigenstate of angular momentum operators $J^2$ and $J_z$.  The state $\psi_0$ is evolved using the kicked top Floquet $U(k)$ in Eq. \ref{eq:uniqkt} and projected onto $\lbrace J_x,\: J_y,\: J_z \rbrace,$ a natural set of observables for the system at each time step. Then, purity at time $m$ is
\begin{equation}
    P_{su(2)}(m;\psi_0,k)= \frac{1}{j^2} \sum_{l=\:x,y,z} |\bra{\psi_0 U^m(k)} J_l \ket{U^m(k) \psi_0}|^2,
\label{eq:ge_purity}
\end{equation} 
where $\frac{1}{j^2}$ is the normalization constant. The corresponding generalized entanglement is given by
\begin{equation}
    \mathrm{GE}_{su(2)}(m;\psi_0,k)= 1- P_{su(2)}(m;\psi_0,k)
\end{equation}
Two distinct periodicities in $k$ are observed in GE as before:

\begin{equation}
 \begin{split}
    GE_{su(2)}(m;\psi_0,&k) = GE_{su(2)}(m;\psi_0, k + 4p\pi j), \;j \in \mathbb{Z}^+\\
           & = GE_{su(2)}(m;\psi_0, k + 2p\pi j), \; j \in \mathbb{Z}^+-\frac{1}{2}
    \end{split}
    \label{eq:ge-period1}
\end{equation}
The period $\kappa_j$ can be identified as $4p\pi j$ and $2p\pi j$ for the two cases in the above equation.
The numerical simulations with kicked top dynamics shown in Fig. \ref{fig:period_GE}(top panel) display the expected periodicity. When $\alpha \neq n \pi/2,$ where $n$ is an integer, the periodicity of $2 \pi j$ or $4\pi j$ is always satisfied, though there can exist periodicity $\kappa_j < 2 \pi j$ or $\kappa_j < 4 \pi j$ consistent with the general rules for $\alpha=n \pi/2.$ Although we chose $su(2)$ due to its physical relevance, it is noteworthy that the periodicity in $k$ is independent of the subspace $\Omega$ one chooses.

\section{Periodicity in Observational entropy}
Observational entropy (OE) is a unification of Gibbs and Boltzmann entropy into one.
Proposed by von Neumann \cite{von2010proof}, now there is a revived interest partly due to its relevance as a signature of chaos \cite{sreeram2023witnessing}.
\begin{figure*}[hbt!]
     \centering
\includegraphics[width=\textwidth]{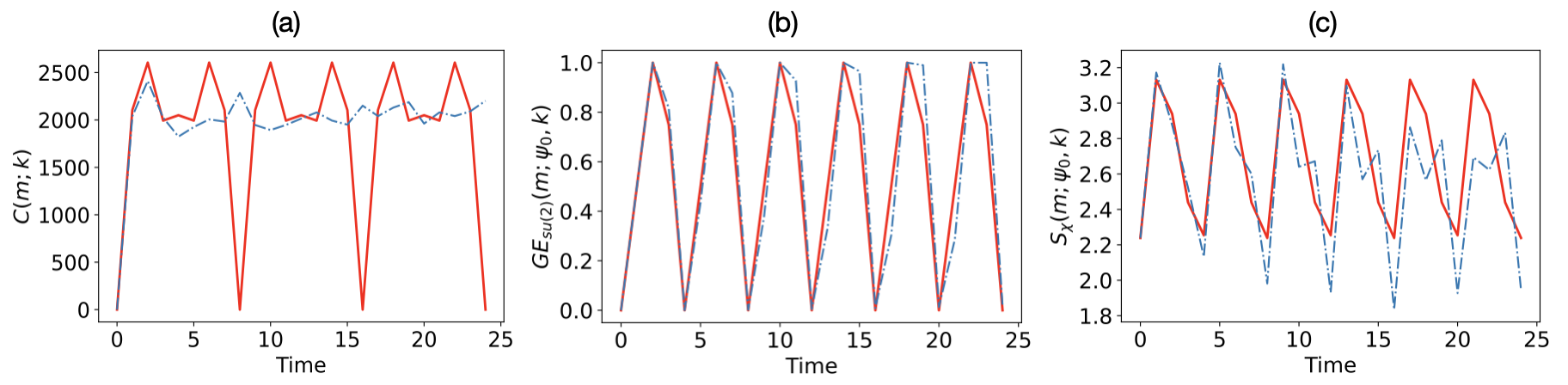}
    \caption{(a) Periodicity of OTOC (b) generalized entanglement and (c) observational entropy for $j=20$, at $k=N \pi/2$, where $N=2j$ is shown in red. The blue dashed line in each figure indicates a slightly different chaoticity, $k=N\pi/2+0.1$ One can see that the periodicity has vanished. For GE and OE, the initial state~($\psi_0$) is a  coherent state at $\theta=\pi/4,\: \phi=\pi/4.$  Periodicity at the special $k$ holds irrespective of the state. }
   \label{fig: period_2jpiby2}
\end{figure*}
Consider a state $\rho$ in $\mathcal{H}.$  
The observational entropy of $\rho$ is defined as:
\begin{equation}
S_\chi (\rho) = -\sum_i p_i \log \frac{p_i }{V_i}. 
\label{eq:oe} 
\end{equation}
where $p_i=\Tr(\Pi_i \rho)$ is the probability to find the state in a macrostate $\mathcal{H}_i$ on projective measurement with the set of projectors $\chi = \lbrace \Pi_i \rbrace.$ The dimension of a macrostate is  $V_i=\Tr(\Pi_i),$ and the macrostates partition the Hilbert space: $\mathcal{H}= \oplus_i \mathcal{H}_i$.
OE can be interpreted as the uncertainty associated with the system state  according to an observer who performs measurements on it. The resolution of the observer determines the value OE takes. 

In practice, it might be easier to have coarse-grained measurements since knowing the complete density matrix would require a large number of measurements. In such cases, an observer can choose a coarse-graining given the physical problem. From a physical perspective, coarse-graining represents a lack of complete knowledge about the system.



 To study periodicity in OE, the initial state is taken to be a random spin coherent state, defined in Eq. \ref{eq:coh}. This state is stroboscopically evolved using the kicked-top unitary $U^m(k)$ for $m$ iterations, and a coarse-grained measurement is performed after each time step. Since we are dealing with the dynamics of OE, we rewrite the notation to illuminate its dependency on the underlying variables as follows.

 \begin{equation}
     S_\chi (m;\psi_0,k) = -\sum_i p^m_i \log \frac{p^m_i }{V_i},
 \end{equation}
 where $p^m_i$ is the probability of observing $i^{th}$ macrostate on measurement at time $m.$

 For the same reasons as for generalized entanglement -- periodic property of  $U^m(k)$ under $k \to k + \kappa_j$ -- OE also shows the same periodicity $\kappa_j$ as obtained in Eq. \ref{eq:ge-period1}. Here, we have used eigenvectors of $J_z$ to partition the Hilbert space, with a constant coarse-graining length of two. Notably, $\kappa_j$ is independent of the coarse-graining. This has been verified by partitioning the Hilbert space differently, and in each case $\kappa_j$ remained unchanged. Figure \ref{fig:period_GE}(bottom panel) shows numerically simulated OE using kicked top unitary and the expected periodicity in OE are indicated by a red vertical line.


\section{Discussion and summary}

The foregoing results show that the dynamical measures display periodicity as a function of kick strength $k$. The existence of periodicity in these measures implies repetitions. Since periodicity $\kappa_j$ depends on $j$, it is clear that the unique behavior of these measures -- such as OTOC -- are limited to $0 \le k \le \kappa_j$. Hence, it is reasonable to identify $k_{\rm max} = 4\pi j$ for integer $j$, and  $k_{\rm max} = 2\pi j$ for $j$ being half-integer as the periods. Thus, $k_{\rm max} \propto j$. Hence, in experimental realizations of kicked top, $k_{\rm max}$ will provide useful guidance for the choice of $j$ such that repetitions of dynamical measures can be avoided. {We  note that this periodicity in the dynamical measures of chaos is not always correlated with recurrences in the quasi-energies. The quasi-energies show the same period of $4\pi j$ in $k,$ as the quantum chaos measures, for integer $j$. However, for half-integer values of $j$, the quasienergy spectrum show a different period than the quantum chaos measures, as explained in the supplementary.}
 

A recent paper \cite{sharma2023violation} showed that the quantum kicked top exhibits integrability at $k=N \pi/2,$ where $N=2j$, while the corresponding classical phase space is chaotic. We examined the dynamical measures of chaos at this special value of $k=k_s=N\pi/2$. It is seen that OTOC, GE and OE exhibit periodic behaviour at $k_s$ even for $j \gg 1$. The numerical simulations shown in Fig. \ref{fig: period_2jpiby2} reveals the periodicity for $j=20$.  This is different from what is anticipated based on Eq. \ref{eq:otoc-period1}. For $j \gg 1$ and the large value of kick strength, one would have expected these measures to grow and saturate in the figure. { Choosing a slightly different $k$ ruins the periodicity as seen in Fig. \ref{fig: period_2jpiby2}}.  It is noteworthy that Loschmidt echo does not show this periodicity. It is because two different chaoticity parameters, $k$ and $k'$, are involved, and the overlap of their dynamics always decays, even if kick strength is chosen to be $k_s=N\pi/2$.

In summary, we have reported the periodicity properties of four different dynamical measures of chaos,
which arises from the inherent periodicity of the kicked top's unitary operator.
The periodicity in $k$ is of experimental significance since this helps in setting the parameter values in experimental platforms to avoid repeating observations. 
Quantum chaos experiments require precise control over the dynamics and need to avoid environmental decoherence. These experiments are costly, and knowing the periodicity would help one design the experiments to avoid repetitions. 
\section{Acknowledgements}
MSS and SPG acknowledge funding from the QuEST program of the Department of Science and Technology~(DST), Govt of India, and IISER Pune, India.


\bibliographystyle{unsrtnat} 
\bibliography{cas-refs}






\end{document}


\title{%
  {\red{Supplementary Information}} \\
  \large Periodicity of dynamical signatures of chaos in quantum
kicked top}

\author{Sreeram PG and M.S Santhanam }
\affiliation{Department of Physics, Indian Institute of Science Education and Research, Pune 411008, India}

\maketitle
\section{The Quantum kicked top}
The time evolution unitary obtained from the kicked top Hamiltonian is given by
\begin{equation}
U= \exp (-i \frac{k}{2j} J_z^2 ) \exp(-i \alpha  J_y).
\end{equation}
This unitary causes precession of the state about the $Y-$ axis by an angle $\alpha$ followed by an impulsive rotation about the $Z-$ axis. Alternatively, in the Heisenberg picture,
the evolution of the angular momentum operator can be obtained as follows:
\begin{equation}
J_i^\prime = U^\dagger J_i U.
\end{equation}
The recursion relations for the angularmomentum components are obtained as follows.
\begin{equation}
\begin{split}
J_x^\prime &= \frac{1}{2} (J_x \cos  \alpha + J_z \sin \alpha +iJ_y)  e^{ i\frac{k}{j}( J_z\cos \alpha- J_x \sin \alpha+\frac{1}{2})}  + \mathrm{h.c.} \\
J_y^\prime &= \frac{1}{2i} (J_x \cos  \alpha + J_z \sin \alpha+iJ_y)  e^{ i\frac{k}{j}( J_z\cos \alpha- J_x \sin \alpha+\frac{1}{2})}  + \mathrm{h.c.} \\
J_z^\prime &= -J_x \sin \alpha +J_z \cos \alpha.
\end{split} 
\label{t}
\end{equation}

The classical limit  can be obtained by taking $j\rightarrow \infty.$  Defining $X= \expval{\frac{J_x}{j}}, ~ Y= \expval{\frac{J_y}{j}} $, and $Z= \expval{\frac{J_z}{j}}$, where the expectation values are  with respect to spin-coherent states, we get the classical map:
\begin{equation}
\begin{split}
X^\prime &= (X \cos  \alpha +Z \sin \alpha)  \cos [k( Z\cos \alpha- X \sin \alpha)] \\ &- Y \sin [k (Z\cos \alpha- X \sin \alpha)],\\
Y^\prime &=  (X \cos  \alpha +Z \sin \alpha)  \sin [k( Z\cos \alpha- X \sin \alpha)] \\ &+ Y \cos [k (Z\cos \alpha- X \sin \alpha)], \\
Z^\prime &= -X \sin \alpha +Z \cos \alpha.
\end{split} \label{top}
\end{equation} 
Taking  $\alpha =\pi/2$,  the above  map  simplifies, and takes the form \cite{haake1987classical}  

\begin{equation}
\begin{split}
X^\prime &= Z \cos (k X) + Y \sin (k X),\\
Y^\prime &= -Z \sin (k X) + Y \cos (k X), \\
Z^\prime &= -X.
\end{split}
\end{equation} 

\begin{figure}[H]
   \centering
    \includegraphics[scale=0.3]{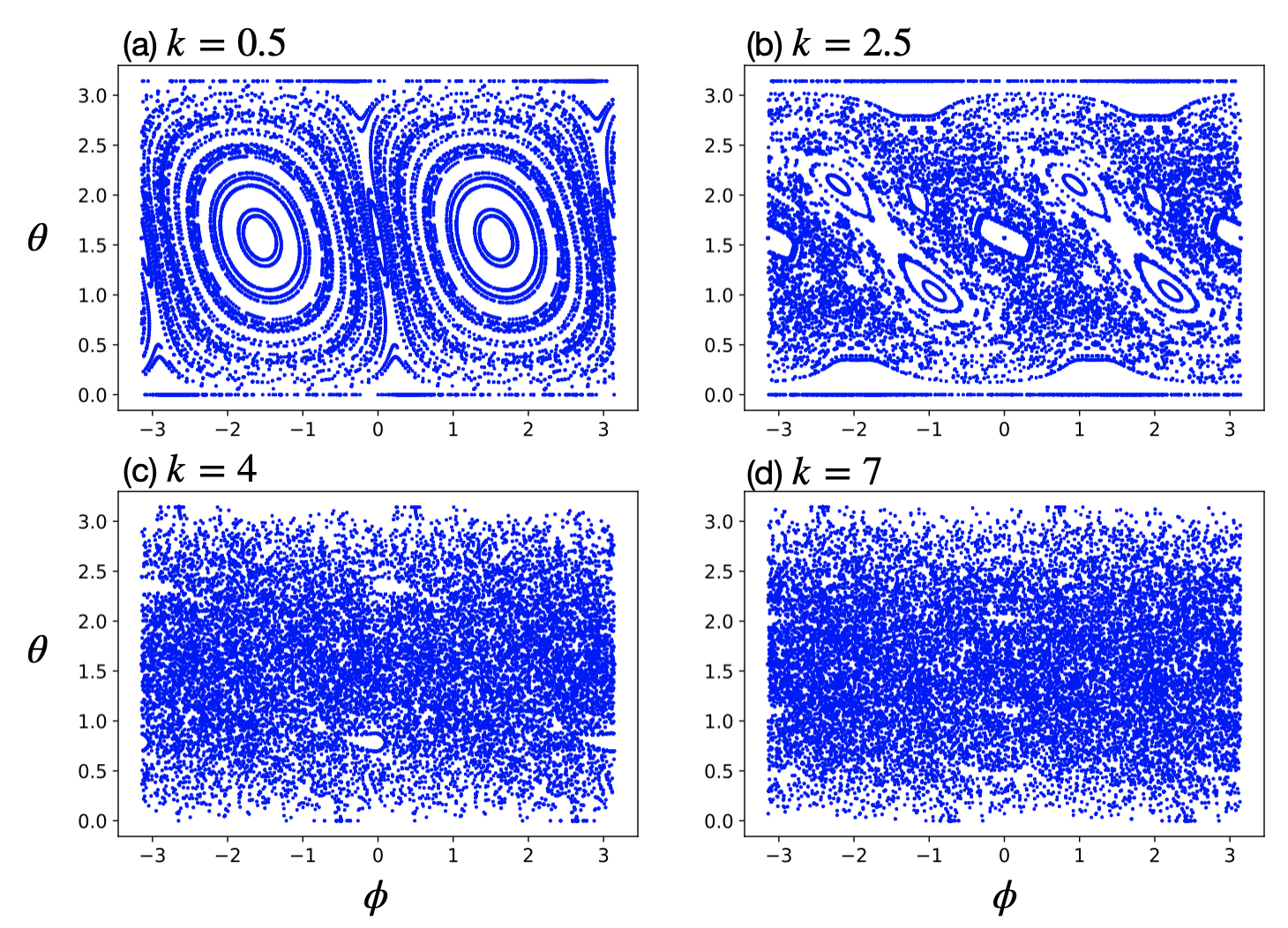}
    \caption{Classical phase space for $\alpha= \pi/2$ at four different chaoticity values, a) $k=0.5$, b) $k=2.5$, c) $k=4$ and d) $k=7.$}
    \label{fig:phasespace} 
\end{figure}

The classical map for the case of $\alpha=\pi/2$, at four different values of $k$ are shown in Fig. \ref{fig:phasespace}. The phase space is mostly regular at $k=0.5.$ As $k$ increases, the regular structures disappear and the phase space becomes increasingly chaotic for $k \gg 1 $. As seen in Fig. \ref{fig:phasespace}(d), the dynamics is predominantly chaotic.

\FloatBarrier

\section{Recurrences in the quasienergies}
For integer values the angular momentum quantum number $j,$ the periodicity found in the chaos measures is $4 \pi j ,$ for the kicked top Floquet given by \begin{equation}
U(k) = \exp (-i \frac{k}{2j} J_z^2 ) \exp(-i \alpha  J_y). 
\label{eq:uniqkt}
\end{equation} We fix $j=2,$ and $\alpha=\pi/4,$ in sync with the demonstrations in the paper, and choose an arbitrary $k,$ the kick strength. Quasi-energies of the Floquet for the first few integer multiples of the period in $k$ is shown in the table below. 
\begin{center}
\begin{tabular}{ |c|c|c| } 
 \hline
 $k=2.1$ &  $k+4\pi j$ & $k+8 \pi j$ \\
 \hline
$-1.2947$ & $-1.2947$  & $-1.2947$ \\
\hline
$-0.2641 $ & $-0.2641$  & $-0.2641$  \\
\hline
0.6478 & $0.6478$&$0.6478$ \\
\hline
0.7806 &0.7806 &0.7806\\
\hline
 1.16346&1.16346&1.16346\\
 \hline
\end{tabular}
\end{center}
Thus, the quasienergies repeat at periodic intervals like the dynamical measures of chaos. We fixed $k=2.1$ above, but the observation is invariant with the choice of $k$.

For half-integer $j$ periodicity in $k$ shown by dynamical chaos measures is that of $2 \pi j$. For $j=1.5,$ and other parameters remaining the same, the quasi-energies for the first two periods are given in the table below.
\begin{center}
\begin{tabular}{ |c|c|c| } 
 \hline
 $k=2.1$ &  $k+2\pi j$ & $k+4 \pi j$ \\
 \hline
$-0.5735$  & -1.3589& $-0.80071$  \\
\hline
$0.7700$  &  -0.0153 &$-0.3910 $  \\
\hline
$1.1797$& 0.3943& -0.1638 \\

\hline
1.4069& 0.6215& 0.9972\\
 \hline
\end{tabular}
\end{center}
The quasienergies are noticeably different. This is because in the Floquet equation at $k+\kappa_j$,
\begin{equation}
\begin{split}
     U({k+ \kappa_j}) & = \exp (-i \frac{k+\kappa_j}{2j} J_z^2 ) ~ \exp(-i \alpha  J_y), \\ 
                         & = \exp(-i \frac{\kappa_j}{2 j} J_z^2)  ~ U(k),
\end{split}
\label{eq:optrans}
\end{equation}
  the unitary prefactor $\exp(-i \frac{\kappa_j}{2 j} J_z^2)= \frac{1}{\sqrt{2}}(1-i) \mathcal{I}_d $ at $\kappa_j= 2\pi j$, and  $\exp(-i \frac{\kappa_j}{2 j} J_z^2)= -i \mathcal{I}_d ,$ at $\kappa_j= 4\pi j$. However, when $\kappa_j= 8\pi j,$ $\exp(-i \frac{\kappa_j}{2 j} J_z^2) = -\mathcal{I}_d $. Therefore, $U(k+8 \pi j)=-U(k)$, and their quasienergies are the same, as listed in the table below.
 \begin{center}
\begin{tabular}{ |c|c| } 
 \hline
  $k=2.1$ &  $k+8\pi j$ \\
  \hline
  -0.5735 & $-0.5735$ \\
  \hline
  0.7700 & 0.7700 \\
  \hline
  1.1797 & 1.1797 \\
  \hline
  1.4069 & 1.4069\\
  \hline
\end{tabular}
\end{center}
Thus, the quasienergies for half-integer $j$ do not show the same period as that of quantum chaos measures discussed, but a higher period, based on the unitary prefactor term. The quantum correlation functions acting as quantum chaos measures are periodic at $\kappa_j=2\pi j,$ and do not require the quasienergies to be the same to show periodicity.

\bibliography{cas-refs}